\documentclass{PoS}
\newcommand{\nn}{\nonumber}
\newcommand{\ovl}[1]{\overline{#1}}

\newcommand{\eqn}[1]{(\ref{#1})}
\newcommand{\p}{\partial}
\newcommand{\tr}{\mathrm{tr}}
\newcommand{\vev}[1]{\left\langle #1 \right\rangle}

\title{1st or 2nd; the order of finite temperature phase transition of
 $N_f=2$ QCD from effective theory analysis}

\ShortTitle{The order of finite temperature phase transition of $N_f=2$
 QCD}

\author{Sinya Aoki\\
Yukawa Institute for Theoretical Physics, Kyoto University,
Kitashirakawa Oiwakecho, Sakyo-ku,
Kyoto 606-8502, Japan\\
E-mail: \email{saoki@yukawa.kyoto-u.ac.jp}}

\author{Hidenori Fukaya\\
        Department of Physics, Osaka University, Toyonaka 560-0043, Japan\\
        E-mail: \email{hfukaya@het.phys.sci.osaka-u.ac.jp}}

\author{\speaker{Yusuke Taniguchi}%
\\
Graduate School of Pure and Applied Sciences,
University of Tsukuba,
 and \\
Center for Computational Sciences, University of Tsukuba,
\\
Tsukuba, Ibaraki 305-8571, Japan\\
        E-mail: \email{tanigchi@het.ph.tsukuba.ac.jp}}

\abstract{
In the previous work, we have shown that the SU(2) chiral symmetry
recovered above the critical temperature gives a strong constraint
on the Dirac eigenvalue spectrum and this constraint is strong
enough for a set of anomalous U(1) chiral symmetry breaking operators to
vanish in the thermodynamical and chiral limits.
We use this condition as an input and impose a constraint on the Landau
low energy effective theory of QCD.
The only constraint we can set is that the mass splitting term between
the pion and eta meson should vanish.
All the singlet/non-singlet scalar/pseudo-scalar mesons contribute to
the effective theory.
We evaluate the renormalization group $\beta$-function for the effective
theory using the $\epsilon$-expansion at one loop level, but find no
stable infra-red fixed point except for the trivial Gaussian one.
The chiral phase transition seems to be of first order.
}

\FullConference{31st International Symposium on Lattice Field Theory
 LATTICE 2013\\
	 July 29 - August 3, 2013\\
	 Mainz, Germany}

\begin{document}

\section{Introduction}

The chiral symmetry $U(N_f)_R\times U(N_f)_L$ is one of the most
important symmetry of the QCD Lagrangian.
This chiral symmetry is known to be broken through two different
mechanisms.
The non-singlet $SU(N_f)_R\times SU(N_f)_L$ part is broken spontaneously
into $SU(N_f)_V$.
On the other hand the singlet $U(1)_R\times U(1)_L$ part is broken
explicitly by the anomaly to $U(1)_V$.

The spontaneously broken non-singlet symmetry is believed to be
recovered above a critical temperature $T_c$.
On the other hand, as the anomaly originates from the cut-off of the
theory, we cannot expect any relation between the $U(1)_A$ and the
restored $SU(N_f)_L\times SU(N_f)_L$ symmetries.
It is natural to assume that the anomalous $U(1)_A$ symmetry remains
broken above $T_c$.

However, in a recent work \cite{Aoki:2012yj}, we have shown that
in the two-flavor lattice QCD, the $SU(2)_L\times SU(2)_R$ chiral
symmetry gives a set of strong constraints on the Dirac eigenvalue
distribution, which has a tight connection to the $U(1)_A$ symmetry.
In fact, the obtained constraints are strong enough for a set of
$U(1)_A$ breaking operators to vanish in the thermodynamical limit
\footnote{
A similar observation was first reported by Cohen
\cite{Cohen:1996ng,Cohen:1997hz}.
Our work has confirmed that the chiral zero-mode's effect, as well as
the effect from the UV-cutoff, he had neglected, do not change his
result.
}.
We have also shown that the $U(1)_A$ breaking $Q=\pm 1$ topological
sector \cite{Lee:1996zy} cannot survive in the infinite volume limit,
and a lattice simulation should have lattice artifacts in the $U(1)_A$
breaking operators (if exists) 
unless it employs the exact chiral symmetric Dirac
operator \cite{Cossu:2013uua}.

The presence/absence of the $U(1)_A$ symmetry at, and above the critical
temperature is important, since it should largely affect the phase
transition \cite{Pisarski:1983ms}.
If the $U(1)_A$ symmetry remains to be broken, the chiral phase
transition is likely to be the second order and its critical behavior
should be in the same universality class as the $O(4)$ sigma-model.
If the $U(1)_A$ is restored at the same critical temperature,
the phase transition is likely to be the first-order
\cite{Pisarski:1983ms}, or at least, be in the different universality
class from $O(4)$, even if it is the second-order \cite{Vicari:2007ma}.

In this work, we investigate the effect of our previous work
\cite{Aoki:2012yj} on the chiral phase transition.
We construct the general effective theory, and give the constraints from
our previous results in the $U(1)_A$ order parameters.
Then, we calculate the $\beta$ functions of the parameters, and study if
there is any infra-red fixed points.
Our result turns out that 
there is no infra-red fixed point at one-loop with the
$\epsilon$-expansion.
If this one-loop observation is correct even at the non-perturbative
level, the chiral phase transition is likely to be the first-order.

\section{Restoration of $U(1)_A$ symmetry}

We briefly review the restoration of the $U(1)_A$ symmetry shown in
Ref.~\cite{Aoki:2012yj}.
We consider the $N_f=2$ QCD and start from an assumption that the
non-singlet $SU(2)_R\times SU(2)_L$ chiral symmetry is fully recovered
above the critical temperature $T_c$.
So any order parameters of the non-singlet chiral symmetry should
vanish.
We consider the following type of the order parameter
\begin{eqnarray}
\frac{1}{V^k}\langle \delta^a{O}_{n_1,n_2,n_3,n_4} \rangle
\end{eqnarray}
given by performing the $SU(2)_A$ transformation $\delta^a$ on a
non-singlet parity odd operator
\begin{eqnarray}
 {O}_{n_1,n_2,n_3,n_4}=
\left(P^a\right)^{n_1}\left(S^a\right)^{n_2}
\left(P^0\right)^{n_3}\left(S^0\right)^{n_4},
\label{eqn:parity-odd-operator}
\end{eqnarray}
where $P^0$ and $P^{a=1,2,3}$ is the singlet/non-singlet pseudo-scalar
operator integrated in the space-time
\begin{eqnarray}
 P^{a}=\int d^4x\ovl{q}\gamma_5\tau^aq(x),\quad
 P^{0}=\int d^4x\ovl{q}\gamma_5q(x)
\end{eqnarray}
and $S^{a=1,2,3}$ and $S^0$ the scalar operator
\begin{eqnarray}
 S^{a}=\int d^4x\ovl{q}\tau^aq(x),\quad
 S^{0}=\int d^4x\ovl{q}q(x).
\end{eqnarray}
A summation is not taken in $a$.
Here we notice that the order parameter should be normalized by an
appropriate power $k$ of the volume since we adopted the integrated
operator.
The value of $k$ is chosen so that the order parameter takes finite
value at zero temperature.

We require all the order parameters to vanish above $T_c$ when the
thermodynamical and then chiral limit is taken in a right way
\begin{eqnarray}
\lim_{m\to 0}\lim_{V\to\infty}\frac{1}{V^k}
\langle \delta^a {O}_{n_1,n_2,n_3,n_4}\rangle =0.
\label{eqn:non-singlet-restoration}
\end{eqnarray}
We convert this requirement to a constraint on the eigenvalue density
(eigenvalue distribution function) $\rho^A(\lambda)$ of the Dirac
operator, where the superscript $A$ intends the gauge configuration
dependence.
We impose a few assumptions on a behavior of the VEV of a mass
independent operator as a function of the quark mass introduced through
the Boltzmann factor.
The eigenvalue density is also assumed to be analytic around $\lambda=0$
\begin{eqnarray}
\rho^A(\lambda)=\sum_{n=0}^{\infty}\rho_n^A\frac{\lambda^n}{n!}.
\end{eqnarray}
Then we have shown that the restoration of the $SU(2)_R\times SU(2)_L$
chiral symmetry \eqn{eqn:non-singlet-restoration} requires the VEV of
the first three coefficients should vanish
\begin{eqnarray}
\lim_{m\to 0}\lim_{V\to\infty}\vev{\rho_0^A}
=\lim_{m\to 0}\lim_{V\to\infty}\vev{\rho_1^A}
=\lim_{m\to 0}\lim_{V\to\infty}\vev{\rho_2^A}
=0.
\label{eqn:rho-vanish}
\end{eqnarray}
Contributions from the exact zero mode were treated independently from
$\rho^A(\lambda)$.
Defining $N_{R+L}(A)=N_R(A)+N_L(A)$ and $Q(A)=N_R(A)-N_L(A)$, where
$N_{L/R}$ is the number of left/right-handed chiral zero-modes, it is
shown that the VEV of these operators should vanish satisfying
\eqn{eqn:non-singlet-restoration}
\begin{eqnarray}
\lim_{V\to\infty}\frac{1}{V^k}\vev{N_{R+L}(A)^k}
=\lim_{V\to\infty}\frac{1}{V^k}\vev{Q(A)^{2k}}
=0.
\label{eqn:zero-mode-vanish}
\end{eqnarray}

These constraints are enough to show the $U(1)_A$ order parameters of
the similar type to vanish above $T_c$
\begin{eqnarray}
\lim_{m\to 0}\lim_{V\to\infty}\frac{1}{V^k}
\langle \delta^0 {O'}_{n_1,n_2,n_3,n_4}\rangle =0,
\label{eqn:singlet-restoration}
\end{eqnarray}
where $\delta^0$ is the singlet $U(1)_A$ transformation and
${O'}_{n_1,n_2,n_3,n_4}$ is an iso-spin singlet parity odd operator
given as a product in \eqn{eqn:parity-odd-operator}.
This means that the anomaly is invisible in the set of VEV with
operators ${O'}_{n_1,n_2,n_3,n_4}$.

\section{Low energy effective theory}

Let us construct the low energy effective theory of $N_f=2$ QCD above
$T_c$ which reproduce the $U(1)_A$ restoration condition
\eqn{eqn:singlet-restoration}.
The corresponding Landau effective theory is given in terms of meson
field
\begin{eqnarray}
\Phi(x)=\frac{1}{2}\left(\sigma(x)+i\eta(x)\right)
+\frac{\tau^a}{2}\left(\delta^a(x)+i\pi^a(x)\right),
\end{eqnarray}
where $\sigma$ and $\delta^a$ denote the scalar singlet and triplet
field respectively, while $\eta$ and $\pi^a$ are pseudoscalar
counterparts.
$\tau^a$ is the Pauli matrix.

The non-singlet ${\rm SU}(2)_L\times$SU$(2)_R$ chiral rotation is given
by $\Phi\to g_L\Phi g_R^{-1}$
where $g_{L/R}\in$SU$(2)$.
The singlet U$(1)_A$ transformation is given by $\Phi\to
e^{i\alpha}\Phi$.
The general form of the renormalizable Lagrangian which is invariant
under SU$(2)_L\times$SU$(2)_R\times$U$(1)_A$ transformation is given by
\begin{eqnarray}
{\cal L}_0=\tr\left(\p_\mu\Phi^\dagger\p_\mu\Phi\right)
+m_\Phi^2\tr\Phi^\dagger\Phi
+\frac{\lambda_1}{2}\left(\tr\Phi^\dagger\Phi\right)^2
+\frac{\lambda_2}{2}\tr\left(\Phi^\dagger\Phi\right)^2.
\end{eqnarray}
An explicit breaking effect due to the anomaly is taken into account by
adding the U$(1)_A$ breaking terms
\begin{eqnarray}
&&
{\cal L}_{{\rm U}(1)_A}={\cal L}_c+{\cal L}_x+{\cal L}_y,
\\&&
{\cal L}_c=c'\left(\det\Phi+\det\Phi^\dagger\right),
\\&&
{\cal L}_x=\frac{x}{4}\left(\tr\Phi^\dagger\Phi\right)
\left(\det\Phi+\det\Phi^\dagger\right),
\quad
{\cal L}_y=\frac{y}{4}
\left(\left(\det\Phi\right)^2+\left(\det\Phi^\dagger\right)^2\right).
\end{eqnarray}
The external source term corresponding to the quark mass may be given by
\begin{eqnarray}
{\cal L}_M=\tr\left(M\Phi\right)+\tr\left(M^\dagger\Phi^\dagger\right),
\end{eqnarray}
where $M$ is a quark mass matrix.

The total Lagrangian
\begin{eqnarray}
 {\cal L}_{\rm tot}={\cal L}_0+{\cal L}_{{\rm U}(1)_A}
\end{eqnarray}
is invariant under SU$(2)_L\times$SU$(2)_R$, parity and charge
conjugation.
${\cal L}_{{\rm U}(1)_A}$ does not break the U$(1)_A$ symmetry
completely.
${\cal L}_c$ and ${\cal L}_x$ are still invariant under $Z_2$ rotation.
${\cal L}_y$ has a $Z_4$ symmetry given by $\alpha=\pi/2$.

\section{Constraints from vanishing order parameters}

In this section, we consider the constraints on the
${\cal L}_{{\rm U}(1)_A}$ under the condition
\eqn{eqn:singlet-restoration}.
The $U(1)_A$ transformation of the operator is given by
\begin{eqnarray}
\delta^0{O}_{n_1,n_2,n_3,n_4} &=&
-2n_1 {O}_{n_1-1,n_2+1,n_3,n_4}
+2n_2 {O}_{n_1+1,n_2-1,n_3,n_4}
\nn\\&&
-2n_3 {O}_{n_1,n_2,n_3-1,n_4+1}
+2n_4 {O}_{n_1,n_2,n_3+1,n_4-1}.
\label{eqn:operators}
\end{eqnarray}

For $n_1+n_2+n_3+n_4=$odd operator the condition
\eqn{eqn:singlet-restoration} is automatically satisfied if we require
the restoration of the non-singlet symmetry by
\eqn{eqn:non-singlet-restoration}.
This is because two equations \eqn{eqn:non-singlet-restoration} and
\eqn{eqn:singlet-restoration} are related by a linear transformation for
$n_1+n_2+n_3+n_4=$odd.
We shall consider $n_1+n_2+n_3+n_4=$even operators in the following.

\subsection{A constraint from $n_1+n_2+n_3+n_4=2$ order parameters}

There are two kinds of order parameters for $n_1+n_2+n_3+n_4=2$ case
given as a difference between the chiral susceptibilities
\begin{eqnarray}
&&
\chi^{\eta-\sigma}=\frac{1}{2V^2}\vev{\delta^0{O}_{0011}}
=\frac{1}{V^2}\left(\vev{{O}_{0020}}-\vev{{O}_{0002}}\right),
\\&&
\chi^{\pi-\delta}=\frac{1}{2V}\vev{\delta^0{O}_{1100}}
=\frac{1}{V}\left(\vev{{O}_{2000}}-\vev{{O}_{0200}}\right).
\end{eqnarray}

The restoration of the $U(1)_A$ symmetry is discussed in terms of the
leading contribution in the thermodynamical limit \cite{Aoki:2012yj}.
From a diagrammatic point of view the leading term in
$\chi^{\eta-\sigma}$ comes from the disconnected diagram of the singlet
scalar field
\begin{eqnarray}
\chi^{\eta-\sigma}=-\left(\frac{1}{V}\vev{S_0}\right)^2
+{\cal O}\left(\frac{1}{V}\right),
\end{eqnarray}
where we used the parity symmetry to eliminate $\vev{P_0}$.
The ${\cal O}(1/V)$ terms are contributions from the connected diagrams.
$\chi^{\eta-\sigma}$ is trivially zero since the simplest order
parameter vanishes above $T_c$
\begin{eqnarray}
\lim_{M\to0}\lim_{V\to\infty}\frac{1}{V}\vev{S_0}=0.
\end{eqnarray}

The second order parameter $\chi^{\pi-\delta}$ is written as a
difference between the pion and the delta meson propagators
\begin{eqnarray}
\chi^{\pi-\delta}&=&
\left.\vev{\pi^b(-p)\pi^b(p)}\right|_{p=0}
-\left.\vev{\delta^b(-p)\delta^b(p)}\right|_{p=0}.
\end{eqnarray}
The U$(1)_A$ breaking ${\cal L}_c$ plays a role of the mass term to
give different masses for each meson
\begin{eqnarray}
{\cal L}_c=\frac{c'}{2}
\left(\sigma^2-\eta^2-\delta^a\delta^a+\pi^a\pi^a\right).
\end{eqnarray}
Tree level propagators of the non-singlet and singlet mesons are given
by
\begin{eqnarray}
&&
\left.\vev{\pi^b(-p)\pi^b(p)}^{(0)}\right|_{p=0}=\frac{1}{m_\Phi^2+c'},
\quad
\left.\vev{\delta^b(-p)\delta^b(p)}^{(0)}\right|_{p=0}=\frac{1}{m_\Phi^2-c'}.
\end{eqnarray}
The condition $\chi^{\pi-\delta}=0$ is satisfied at tree level if we set
a constraint $c'=0$.

A general form of the two point function including the quantum correction
is given by
\begin{eqnarray}
&&
\left.\vev{\pi^b\pi^b}\right|_{p=0}
=\frac{1}{m_\Phi^2+\delta m_\Phi^2 +c'+\delta c'}
,\quad
\left.\vev{\delta^b\delta^b}\right|_{p=0}
=\frac{1}{m_\Phi^2+\delta m_\Phi^2 -\left(c'+\delta c'\right)}.
\end{eqnarray}

The condition $\chi^{\pi-\delta}=0$ is satisfied by setting a constraint
on the renormalized parameter $c'_{\rm R}\propto c'+\delta c'=0$.
We may consider two possibilities to accomplish this condition.
One is to fine tune the bare parameter $c'$.
The other is to protect $c'_{\rm R}=0$ by a symmetry.
If we set $c'=x=0$ the remaining Lagrangian ${\cal L}_0+{\cal L}_y$
acquires a $Z_4\subset$U$(1)_A$  symmetry.
$c'_{\rm R}=0$ is guaranteed at any loop level and the order parameter
vanishes automatically.

\subsection{Constraint from general $n_1+n_2+n_3+n_4$ order parameter}

We consider order parameters with general $n_1+n_2+n_3+n_4=$even
operators.
However as we shall see shortly the restoration condition with them does
not give any further constraint than $c'_{\rm R}=0$.
This is mainly due to a fact that our constraint
\eqn{eqn:singlet-restoration} is given with the appropriate volume
normalization factor $V^k$ and disconnected diagrams tend to dominate.

The general order parameter is given by four terms in
\eqn{eqn:operators}.
The leading terms in the order parameter can be categorized into three
types from a diagrammatic point of view.
\begin{enumerate}
 \item All the leading terms contain the singlet scalar operator $S_0$.
       In this case all the leading term should be proportional to the
       vev of singlet scalar field and vanish automatically above
       $T_c$.
 \item Only a part of the leading terms contain the singlet scalar
       operator $S_0$.
       This happens for the operator ${O}_{2k_1+1,2k_2+1,2k_3+2,0}$.
       One can easily see that the leading term is proportional to
       either the singlet scalar $S_0$ or a difference of the
       operator $P^aP^a-S^aS^a$.
       The disconnected contribution from the latter should vanish as
       $\chi^{\pi-\delta}=0$ is satisfied by $c'_R=0$.
 \item Any leading term does not contain the singlet scalar operator.
       This is the case for ${O}_{2k_1+1,2k_2+1,0,0}$.
       The leading term of this operator is proportional to
       $P^aP^a-S^aS^a$ and vanishes as in the case of 2 above.
\end{enumerate}

The restoration of the non-singlet $SU(2)_L\times SU(2)_R$ symmetry and
the constraint $c'_{\rm R}=0$ is enough to eliminate all the $U(1)_A$
order parameters of the form \eqn{eqn:singlet-restoration}.
We notice that the $U(1)_A$ breaking interaction term with parameters
$x$, $y$ cannot be constrained at all.

\section{Renormalization group analysis of phase transition}

The last step is the renormalization group analysis of the low energy
effective theory according to Ref.~\cite{Pisarski:1983ms}.
Our effective Lagrangian above $T_c$ is given by
\begin{eqnarray}
{\cal L}_{\rm tot}={\cal L}_0+{\cal L}_x+{\cal L}_y.
\end{eqnarray}

We start by deriving the renormalization group function of the four
couplings $g_i=\left\{\lambda_1,\lambda_2,x,y\right\}$.
The $\beta$-function is defined in a standard manner by taking a
derivative of the renormalized coupling
\begin{eqnarray}
\beta_{g_i}\left(\lambda_1,\lambda_2,x,y\right)=\mu\frac{\p}{\p\mu}g_i(\mu),
\end{eqnarray}
where we adopted the MS scheme with the dimensional regularization
$d=4-\epsilon$.

We shall evaluate the $\beta$-function using the $\epsilon$-expansion at
one loop level in this proceedings.
The $\beta$ function for each coupling turns out to be
\begin{eqnarray}
&&
\beta_{\lambda_1}=-\epsilon\lambda_1
+\frac{1}{8\pi^2}\left(8\lambda_1^2+8\lambda_1\lambda_2+3\lambda_2^2
+\frac{3}{2}x^2+\frac{5}{4}y^2\right),
\\&&
\beta_{\lambda_2}=-\epsilon\lambda_2
+\frac{1}{8\pi^2}\left(4\lambda_2^2+6\lambda_1\lambda_2
-\frac{3}{4}x^2-y^2\right),
\\&&
\beta_{x}=-\epsilon x
+\frac{1}{8\pi^2}x\left(12\lambda_1+6\lambda_2+3y\right),
\\&&
\beta_{y}=-\epsilon y
+\frac{1}{8\pi^2}\left(6\lambda_1y+\frac{3}{2}x^2\right).
\end{eqnarray}

We search for the fixed points to satisfy
$\beta_{\lambda_1}=\beta_{\lambda_2}=\beta_{x}=\beta_{y}=0$ staying on
the critical surface $m_{\Phi R}=c'_R=0$, which is accomplished by fine
tuning the bare parameters $m_\Phi$ and $c'$.
We found six fixed points.
Only one of them at $\lambda_1=\lambda_2=x=y=0$ is the stable ultra
violet (Gaussian) fixed point.
Remaining five fixed points are saddle point and are not stable.
We do not find any stable infra red fixed points for our low energy
effective theory at one loop.
Our perturbative analysis shows that the chiral restoration phase
transition is likely to be of first order.

\section{Discussion}

In the previous work we have shown that the $U(1)_A$ order parameters
written in terms of the scalar/pseudo-scalar operators should vanish above
$T_c$ for the $N_f=2$ QCD as the thermodynamical and the chiral limit is
taken in the right way.
We use this condition as an input and impose a constraint on the low
energy effective theory.
The only constraint we can set is that the mass splitting term between
the pion and eta meson should vanish as $c_R'=0$.
All the singlet/non-singlet scalar/pseudo-scalar mesons $\pi$, $\eta$,
$\sigma$, $\delta$  contribute to the effective theory.
However the $U(1)_A$ symmetry breaking interaction term still remains
with the coefficient $x$ and $y$.

We evaluate the $\beta$-function of the four couplings in the effective
theory at one loop level and search for the stable infra-red fixed point
on the critical surface $m_{\Phi R}=c'_R=0$.
We cannot find any such fixed points except for the trivial Gaussian
fixed point.
The chiral phase transition seems to be of first order.

However the $\epsilon$-expansion may not be sufficient to have a rigid
conclusion.
We need to set $\epsilon\to1$ in the end to discuss the second order
phase transition.
Non-trivial fixed points, which are proportional to $\epsilon$, tend to
be in a strong coupling region.
A higher loop \cite{Vicari:2007ma} or non-perturbative analysis may be
required, with which one may find a stable IR fixed point
\cite{Vicari:2007ma}.
In this case a position of the fixed point would be important.
If a stable IRFP was on $x=0$ the low energy theory would acquire higher
symmetry $SU(2)_L\times SU(2)_R\times Z_4$ on the fixed point.
The universality class would be different from the widely believed
$O(4)$.

This work is supported in part by  the Grant-in-Aid of the Japanese
Ministry of Education (No. 22540265, 25287046, 25800147), 
the Grant-in-Aid for Scientific Research on Innovative Areas 
(No. 2004: 20105001, 20105003, 23105710,  23105701) and SPIRE
(Strategic Program for Innovative Research).

\end{document}